# Role of Artificial Intelligence, Clinicians & Policymakers in Clinical Decision Making: A Systems Viewpoint

1st Avishek Choudhury,MS
*School of Systems and Enterprises*
*Stevens Institute of Technology*
Hoboken, USA
achoudh7@stevens.edu

2nd Onur Asan,PhD
*School of Systems and Enterprises*
*Stevens Institute of Technology*
Hoboken, USA
oasan@stevens.edu

3rd Mo Mansouri,PhD
*School of Systems and Enterprises*
*Stevens Institute of Technology*
Hoboken, USA
mo.mansouri@stevens.edu

*Abstract*—"What is a system?" Is one of those questions that is yet not clear to most individuals in this world. A system is an assemblage of interacting, interrelated and interdependent components forming a complex and integrated whole with an unambiguous and common goal. This paper emphasizes on the fact that all components of a complex system are interrelated and interdependent in some way and the behavior of that system depends on these interdependencies. A health care system as portrayed in this article is widespread and complex. This encompasses not only "hospitals" but also governing bodies like the FDA, technologies such as AI, biomedical devices, Cloud computing and many more. The interactions between all these components govern the behavior and existence of the overall healthcare system. In this paper, we focus on the interaction of artificial intelligence, care providers and policymakers and analyze using systems thinking approach, their impact on clinical decision making.

## I. INTRODUCTION

Being a complex and dynamic industry, Healthcare is shaped by several interacting forces ("Fig. 1Forces shaping clinical judgmentsfigure.1") that contribute to clinical decision making. Clinical decision making is a structural progression, iterating within clinician's mind which is not only guided by experiences but also technology, regulations, physical environment, human factors, and past events. Since the healthcare industry shares cohesions with industrial sectors embracing vulnerability to human cognitive errors, [1] a better indulgent into the evidence on cognitive biases shaping medical decisions is crucial. Such indulgence is predominantly needed for clinicians, as their flawed decisions can be fatal and expensive. In the recent past, we acknowledged the importance of patientlevel and hospital-level factors coupled with medical errors. For instance, standardized approaches like Advanced Trauma Life Support and ABCs for cardiopulmonary resuscitation, at the healthcare system level decreased medical errors [2]. Unfortunately, physician-level factors were overlooked [3]. The presence of latent biases within clinicians have been largely neglected by healthcare management and researchers.

However, recently cognitive biases have been recognized as one of the potent influencers to flawed clinical decisions. We define cognitive biases as "latent distortions in judgment".

According to The Joint Commission, cognitive biases have led to a number of custodian events, like wrong-site surgeries (e.g., confirmation bias), patient falls (e.g., availability heuristic and ascertainment bias), delayed treatment, and diagnostic errors (e.g., anchoring, availability heuristic, framing effect and premature closure) [4]. Traditional decision making was primarily experienced based. Even a clinical decision by a trained professional was steered by a plethora of prior "organizational" decisions. For instance, a surgeon's preference to a given technique for certain operations are conditioned by prior experiences, such as types of operating rooms, equipment, the quality of assisting staff, and the schedule itself [5]. In other words, "micro" decisions concerning individual's clinical judgment and "macro" decisions encompassing organizationwide policy issues are highly inter-depended. The nature of clinician's contribution to clinical decision making must be implied within this context.

However, lately, artificial intelligence (AI) has gained a significant control on clinical decision making. In medicine, AI research is burgeoning rapidly, allowing computers to make data driven predictions [6]. AI has been extensively used to predict and diagnose diseases [7] and has been demonstrating excellent results, specifically in detecting cancer using image classification [8]. In February 14, 2018 — The U.S. Food and Drug Administration (FDA) gave marketing clearance for "Viz.AI's Contact" [9] application that uses artificial intelligence to analyze CT images for stroke indicators.

The performance of AI from a data science perspective has been very promising; Unfortunately, its clinical value is yet to be realized. The complicated and opaque structure of the machine learning algorithms (deep-learning) limits the understanding of this technology and its reliability in the healthcare domain. This has also led to increasing concern regarding its medico-legal and ethical impact [10], [11]. Despite the lack of understanding of AI algorithms, recently, several algorithms have earned regulatory approval for clinical use, and the barricade for entry of novel advanced algorithms has precisely AI on clinical practices and FDA regulations; How it shifted the healthcare paradigm.

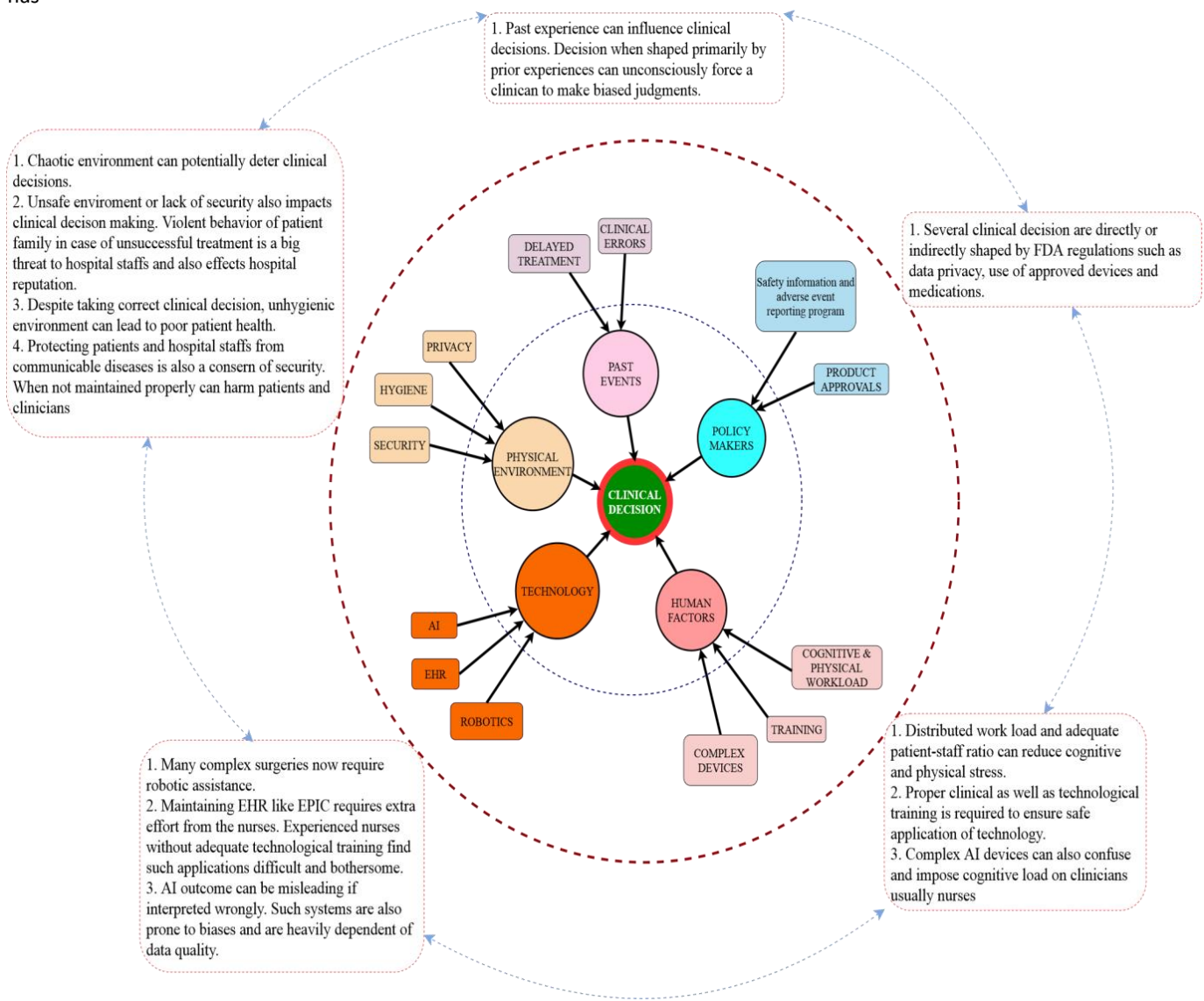

Fig. 1. Forces shaping clinical judgments.

been low [12]. To ensure prudent clinical decisions, regulatory authorities should confirm that AI algorithms meet accepted standards of clinical benefit, just as they do for clinical therapeutics and predictive biomarkers.

In this paper, we adopted a systems perspective to briefly analyze the role of AI in clinical decision making. Consecutively, providing with a brief framework guiding clinical professionals and regulatory bodies like FDA to critically evaluate feedbacks within the healthcare system and thus address concerns regarding clinical judgments. Although not exhaustive, these criteria can ensure prudent clinical decisions. Feedback is the conduction and return of information. It provides evidence to the system that lets it know about its performance relative to a preferred state. To do so, we conceptually analyze the influence of technology

## II. THE BIGGER PICTURE

A typical healthcare system consists of subsystems like "Patient", "Payor", "Provider" and "Policymaker" ("Fig. 2Interaction between sub-systemsfigure.2"). The fundamental goal of these subsystems is equifinal. In other words, each subsystem has its own route to achieve the fundamental goal of maintaining quality of care or safety, which highly depends on correct and timely clinical judgment. Like any other decision, an ideal clinical decision should be logical, rational, unbiased, educated and for the benefit of patient's health. Clinical decision making, although not limited to, but is generally dissuaded by insufficient information, and misinterpretation of health information [13]. However, with the increasing involvement of complex AI systems assisted by availability biases" among clinicians were responsible for 77% of diagnostic inaccuracies in case-scenarios [14]. Out of 2 studies assessing the impact of cognitive biases on clinical decision, only one advocated that higher tolerance to indistinctness was concomitant with medical complications [14].

A lot has been done to compensate for human errors in clinical judgment. Policymakers and other regulatory bodies have developed standards for medical devices and protocols for treatments that would assist clinicians in structuring their decisions. Moreover, due to the complexity of the healthcare system and all the forces shaping this industry, it is arduous to isolate and analyze any subsystem and determine its influence on clinical decisions. However, all subsystems (patients, human factors in healthcare, the effect of cognitive load, cognitive biases, and disrupted communication have also been identified as significant shaping forces steering clinical decisions.

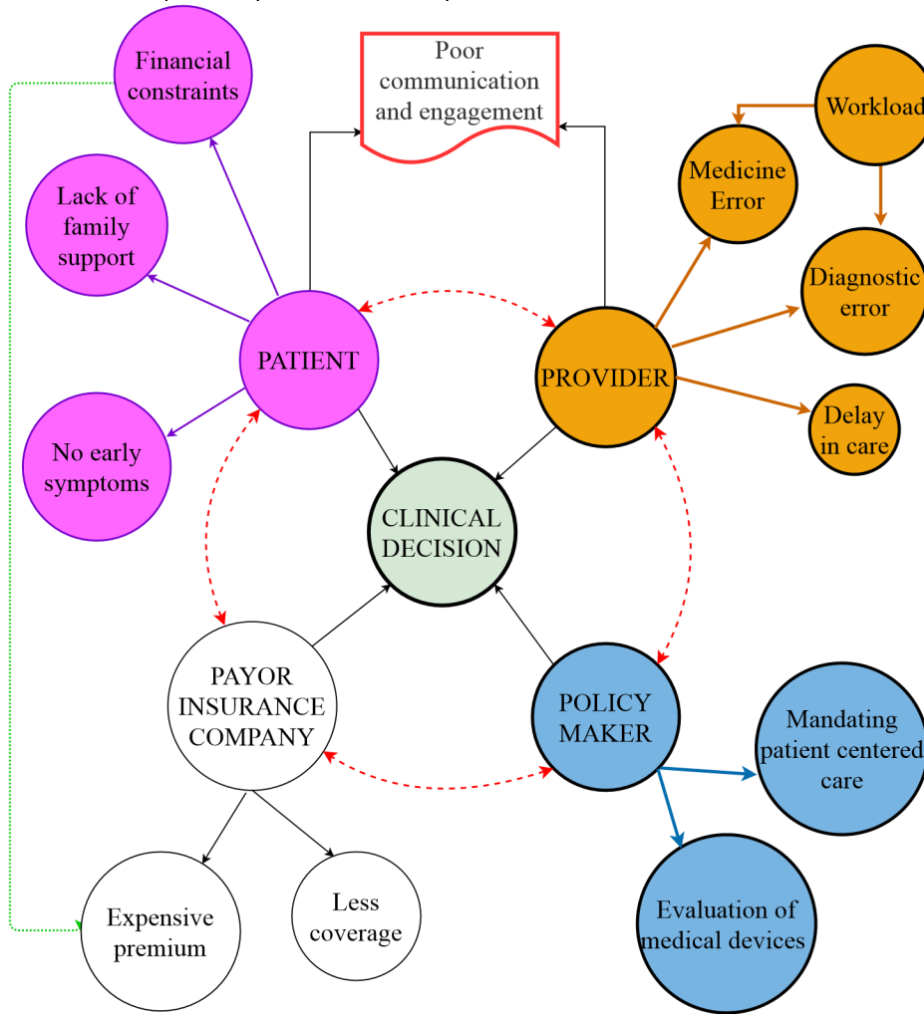

Fig. 2. Interaction between sub-systems.

There are about 19 cognitive biases that impede clinical decision making resulting in an inaccurate diagnosis. Studies have shown that cognitive biases like "overconfidence," "risk averseness," "the anchoring effect," and "information and providers, payers, and policy makers) use AI or other technology to shape their respective decisions. Even though with a unique approach, people use AI to improve overall health of a patient. AI being the common tool, we discuss its interaction with the providers and policymakers.

## III. INVASION OF THE ARTIFICIAL INTELLIGENCE

Clinical decision support systems (CDSS) are commonly used in medicine and have credible impact providing assistance on the safe prescription of medicines, [6] disease diagnosis, [11] and risk screening [6]. These systems are based on predefined rules which are easy to understand and usually shown to minimize clinical errors. However, with increasing size and complexity of health data, these simple rule-based systems fail. As a result of this gap, majority of researches are focusing on developing sophisticated machine learning algorithms to handle complex and big data [6].

Machine learning algorithms have been focusing on a clinical decision support system, often in specific domain like automating triage process [12] and prioritizing individual's access to healthcare services by screening referrals [6]. Such a complex AI system usually require technical knowledge and domain expertise to operate thus it imposes cognitive load on the user because most clinicians especially nurses are not trained to understand and use AI technology. Unfortunately, this deterring effect has been neglected so far. These systems also has the latent potential to entail ethical issues by embedding inequality, consonant to those seen in the automation of job applicant screening, of which it is said that "blind confidence in automated e-recruitment systems could have a high societal cost, jeopardizing the right of individuals radiology. Additionally, machine learning algorithms are also being used as a risk detection system to identify sepsis, mesothelioma, [11] hospital readmissions [15] and others. The performance of machine learning algorithms relies on the precise composition of their training (input) data and training parameter. Even regulating these parameters, some machine learning algorithms fail to produce identical results which might affect clinical decisions [6], [12].

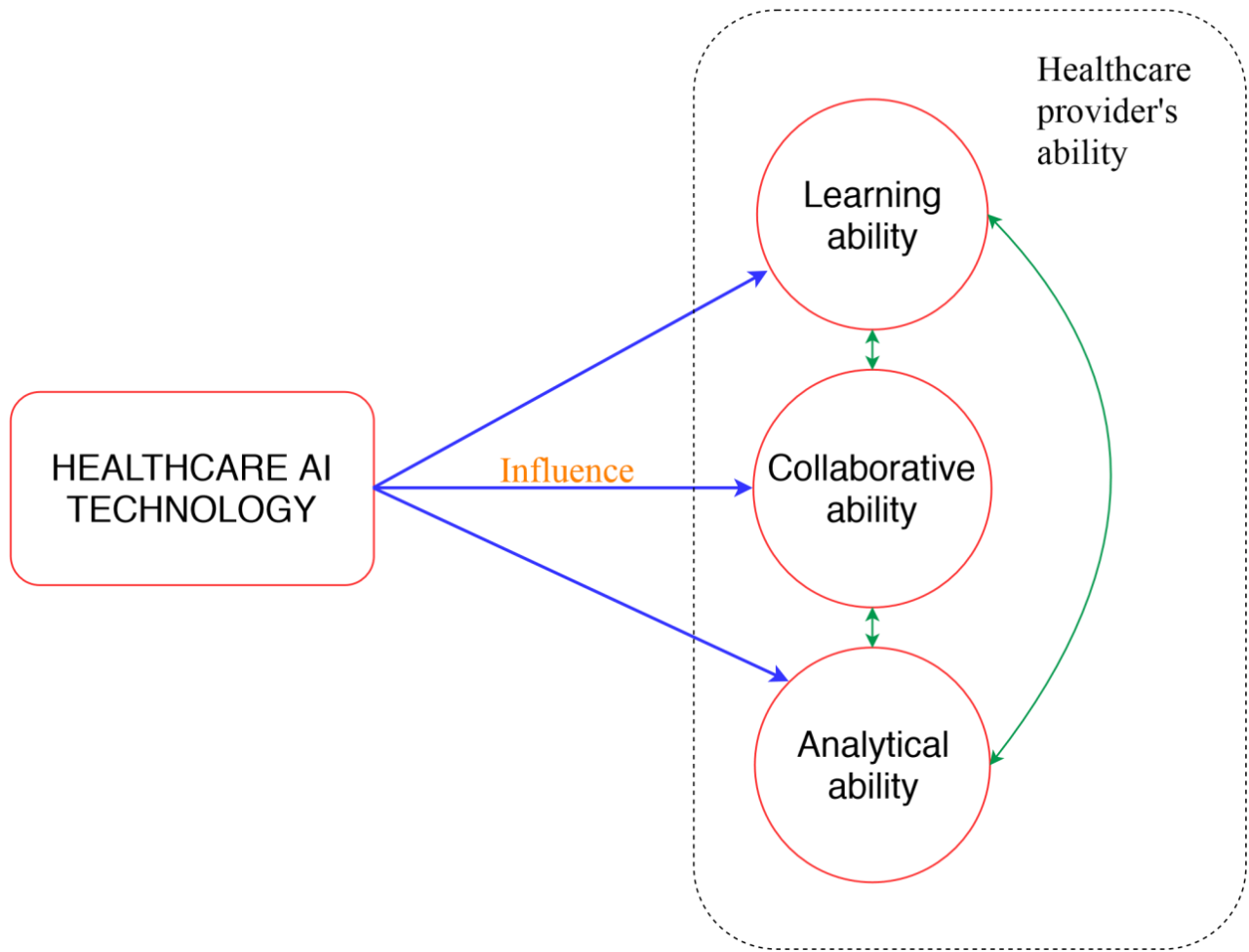

Fig. 3. Interaction between AI and healthcare providers.

Recently AI algorithms are developing systems that go beyond predicting clinical actions. Such advanced systems are to equal opportunities in the job market" [6]. However, this is a complex topic and beyond the remit of this analysis.

## IV. ARTIFICIAL INTELLIGENCE AND THE "PROVIDERS"

AI application continues to develop within the healthcare domain. In the healthcare communication process all participants such as device manufacturer, pharmacy network, inpatient/outpatient clinics, nurses and doctors, etc., are using AI to assist in areas which they are currently struggling, such as in managing patient and staff network, communicating with partner clinics, managing inventory and streamlining professional communication within all providers ("Fig. 3Interaction between AI and healthcare providersfigure.3").

### A. Learning ability

Clinicians need intelligent decisions to manage complex patients and dynamic healthcare department. However, no single model has the ability to describe the healthcare system and its network characteristics. Fortunately, AI has entered the cognitive age, and via deep learning, computers can use existing data to gain insight into the healthcare process, understand system's traffic, and predict outcomes with some certainty.

Due to the dynamic nature of healthcare, the healthcare management can only understand the local state information (such as a single department within the hospital) without the knowledge about the greater system, its internal state and interdepartmental interactions.

AI systems happen to have the capability to deal with fuzzy logic and uncertainty reasoning. AI algorithm such as neural network need not explain the mathematical model of the entire healthcare system but can deal with uncertainty or even unknowability.

### B. Collaborative ability

Adoption of AI-enabled systems by healthcare and insurance providers have expanded the communication network both in scale and size, the structure of communication is thus burgeoning rapidly. Thus, we need multi-agent collaboration of distributed AI into the healthcare network to provide us with real-time information of the communication network and also establish strong security protection and behavioral analysis to identify data breaching activities. This shall enhance the ability to collaborate between healthcare stockholders and streamline healthcare process.

### C. Analytical ability

The healthcare industry previously has produced large amounts of data, driven by record keeping, compliance and regulatory necessities, and patient care [16]. Unfortunately, the majority of the data is stored in hard copy form. Digitizing and structuring (data preparation) can be expensive and timeconsuming.

AI brings along with it the computing power, that is, the ability to analyze the big data that has been generated by the healthcare industry. Industries have started training AI algorithms on these historic data to promote evidence-based medicine, patient profile analysis, disease diagnosis, genomic analysis and others without ensuring data quality.

Although there is a vast amount of data and researches claims that these data can enhance machine learning performance [16] thus strengthening CDSS, there is no guideline to validate the quality and measure the diversity of the data. Any intrinsically embedded bias in the data can skew the final result.

To minimize latent risks associated with big data in healthcare, clinicians and regulatory bodies like FDA should ensure variety, velocity and volume of health-related data. However, the velocity and volume are taken care off, the already intimidating volume of existing healthcare data incorporates personal medical records, radiology images, clinical trial data from FDA submissions, human genetics and population data genomic sequences, [16] and others.

## V. ARTIFICIAL INTELLIGENCE AND THE "POLICY MAKER"

AI and augmented computing capabilities have long held the promise of refining prediction and prognostication of diseases and events in health care. Recently, several algorithms have achieved regulatory approval for clinical use. To unlock the potential of predictive analytics, regulatory authorities should ensure that proposed algorithms meet accepted standards of clinical benefit. External validation and rigorous testing of algorithms are mandatory requirement, [17] but regulatory consents raise apprehensions over the rigor of this process. Given these concerns, we suggest standards to guide the regulation of devices based on predictive analytics just like the TRIPOD Checklist [18].

Many researches focus on the application of AI in healthcare believing to develop a better decision support system [19], [20]. However, unlike a drug or medical device, algorithms inherit uncertainty. The performance of any AI-based algorithm relies on thousands of variables that fluctuates with context. The predictive performance and robustness certainly will alter with a change in the quality as well as the quantity of the input data [11]. Thus, we identified 5 characteristics ("Fig. 4Role of policy makers in safeguarding healthcare AI systemsfigure.4") that regulatory bodies like FDA should address to safeguard and improve healthcare AI systems.

### A. AI outcome

Most assessments of algorithms in medicine domain evaluates AI performance using abstract measures such as area under the receiver operating characteristic curve or classification accuracy [11], [15], [20]. Such metrics are neither easily understandable by clinicians nor by patients. They are time-consuming and typical are not clinically meaningful. Moreover, it does not screen for any biases in the input data.

Future methods of evaluation should consider using established standards of clinical benefit including downstream results such as overall survival but should also incorporate other relevant performance measures such as positive predictive value, number of misdiagnoses, and further diagnostic test characteristics such as false positive and false negative. As it does for drug approvals, the FDA should rigorously confirm and test surrogate endpoints in potential evaluations of advanced algorithms, to prohibit the introduction of questionable algorithms to the medical field where human life is at risk.

Unfortunately, limited algorithms receiving regulatory clearance yield desirable outcomes. As governing bodies decide on which downstream results matter for predictive algorithms, they should also keep in mind that predictive algorithms based on subjective clinician data, or outcomes that rely on access to health care—could systematically inherit bias towards certain clusters of patients [6]. Clinicians' responses to such biased outputs could extend existing bias— and possibly harm patients.

*B. AI standard*

Products developed on predictive algorithms are rarely evaluated against a standard of care. When benchmarking an AI model, its fragility and uncertain behavior has to be considered. For instance, an algorithm that is trained on autism data may classify autism disorder [20] more precisely than developers to design detailed specification of variable data and ensure that predictive algorithms achieve replicable and effective results across institutions.

The FDA announced recently that it is developing a framework for regulating AI products that self-update based on new data. While FDA authorized AI-based devices for detecting diabetic retinopathy and for alerting providers of a potential stroke in patients last year, these products utilize "locked" algorithms that do not continually adapt or learn every time the algorithm is used [21]. The agency is thus seeking comments on how to regulate self-updating algorithms, issuing a new white paper that outlines its initial thoughts.

*D. Compliance to legislation*

AI systems are influencing many process ad patient health within healthcare industry. Along with clinical legal aspects

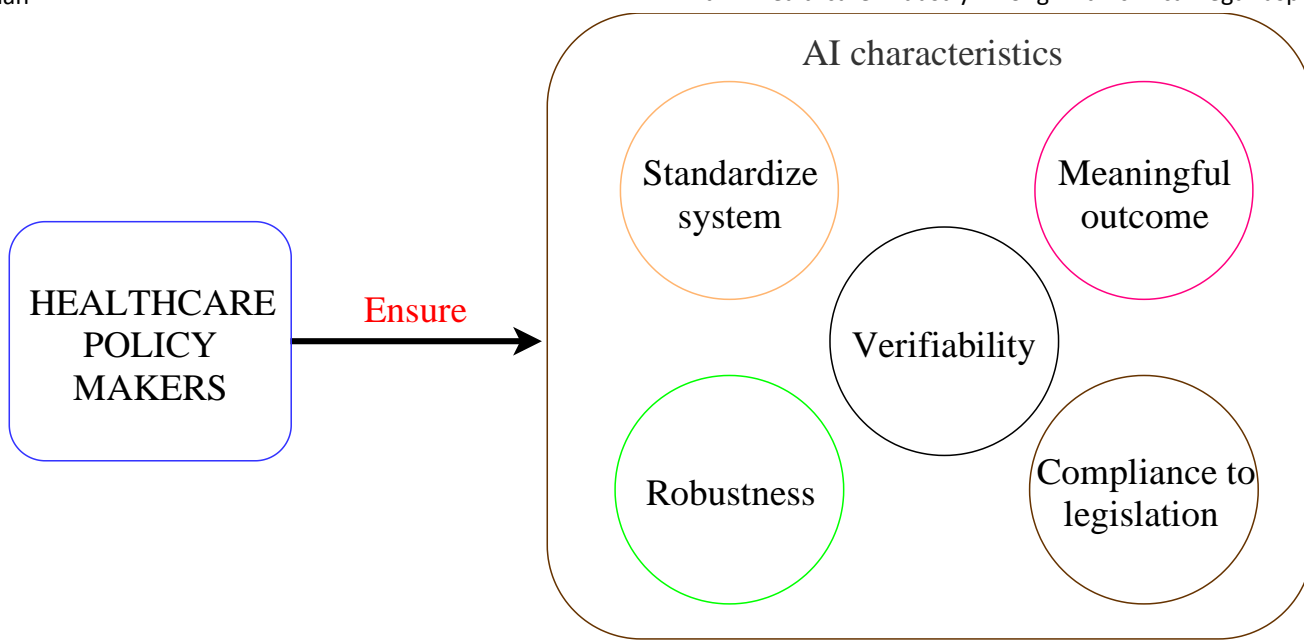

Fig. 4. Role of policy makers in safeguarding healthcare AI systems.

physicians, and thus will give the impression of superiority [11].

Algorithm not trained on variety of data set might result in miss-classification. Thus, it is important to evaluate AI algorithm's output against clinicians' expert and informed judgment.

*C. Interoperability and generalizable property*

Algorithms with FDA 510(k) clearance are available for broad use. Unlike the WAVE platform, other machine learning algorithms, particularly those based on institution-specific EHR or image datasets are typically not be translatable across other EHRs.

In a different clinical setting, interoperability problems and unfamiliarity with the user interface may introduce cognitive disturbances and thus incumber clinicians' capacity to respond to a predictive algorithm output. This necessitates algorithm such as proper task assignment and adherence to protocols, an assignment of responsibility when AI system goes wrong should be developed. To develop such regulation and law, AI will have to become explainable.

Since an algorithm's systemic bias against certain groups may burgeon when deployed across large and diverse populations and its predictive performance may alter based on the quality (variety) of data. AI algorithms should be subject to audits using similar approach as of the FDA Sentinel program for approved drugs and devices after FDA clearance [12].

*E. Verifiable AI system*

The black-box nature of AI systems effects people's trust on the technology. For instance, in healthcare the use of AI systems which can be interpreted and scrutinized by doctors is an absolute requirement. A study reports an incident where an AI system gave a wrong conclusion while predicting pneumonia [22]. This black-box model due to systematically biased data learned that asthmatic patients with heart problems have a lower risk of dying due to pneumonia than a healthy individual. A clinician would realize that this cannot be true as asthma and heart problems are factors which deters prognosis for recovery.

## VI. THE 3 "I" WITHIN HEALTHCARE SYSTEM

From systems thinking perspective, the complexity and behavior of a decision support system with the healthcare domain can be described by 3 characteristics: Interdependences, Interactions and Inter-relationships within the system. "Fig. 5Causal loop diagramfigure.5" although not exhaustive, shows the relationships among most of the major factors within the healthcare industry that influences clinical

### A. Sub-optimization

Sub-optimization refers to the tradition of converging on one component of a system and making changes intended to improve it in isolation. This approach ignores the effects of isolated optimization on the overall system.

Within the healthcare domain, even though clinicians and other non-clinical staffs are highly trained, skilled and bounded by confidentiality and ethics, we have not yet realized that just like AI algorithms, humans are mostly unaware of their biases. One of the main factors that determines the quality of clinical decision is the magnitude of awareness of cognitive bias and data bias among the clinicians.

Artificial intelligence is crunching through medical data, assisting in image recognition and even suggesting the decision support system. As indicated, application of AI, human error, prior knowledge, information quality, cognitive biases, and bias in data influences clinical decision making. An intuitive and typical behavior given aforementioned circumstances is to maximize all the influencers to optimize the fundamental goal of taking prudent clinical decisions. Using "Fig. 5Causal loop diagramfigure.5" as an example, we address the consequence of sub-optimization within healthcare.

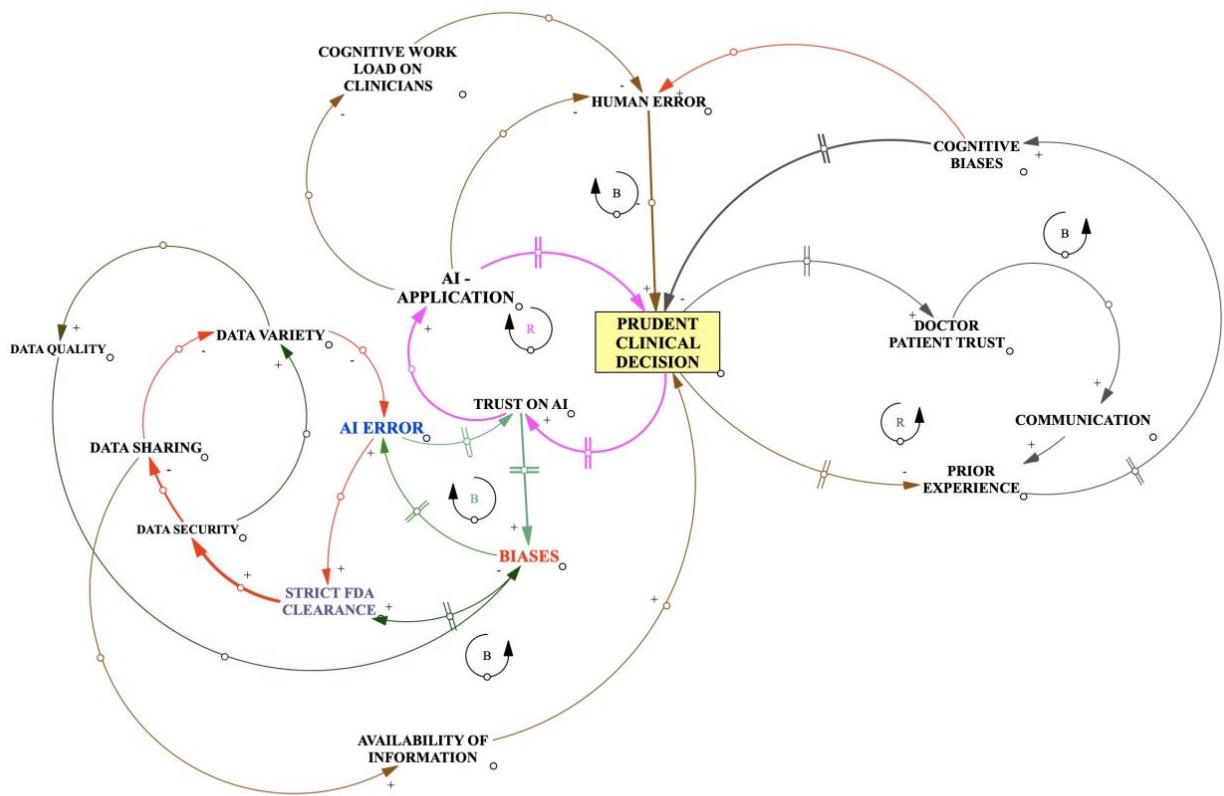

Fig. 5. Causal loop diagram.

diagnosis. But the application of such technology in healthcare has associated concerns regarding data bias. That being said, access to variety and high volume of data is one of the fundamental prerequisites for developing an unbiased machine learning algorithm; However, as access to data increases, the potential threat to data privacy and security increases. Thus, sub-optimization of either data sharing (unbiasing AI algorithms) or data privacy can deter the overall healthcare system. Similarly, if the system converges to maximize trust on AI, then biases (automation bias) will be induced among the clinicians and the patients. Even though, it will encourage the application of AI and reduce work load, the overall quality of clinical decision will be compromised.

Thus, sub-optimization embeds chaos into the system. To attain homeostasis, healthcare system should adapt to the immediate and delayed feedback from all interacting components.

## VII. CONCLUSION

In this paper, we define healthcare system as an integration of other systems and discussed how the intersection of systems like FDA, Computing technology, Clinicians (human system), etc., governs the overall healthcare industry and manipulates clinical decision system. A metaphorical example to this can be the interaction and dependency between a computer system (device) and the internet, both of which are independent systems with respective features; However, their value, and behavior changes at their intersection. The internet system enhanced the ability of a computer system but also exposed it to threats such as viruses and data theft. To mitigate and prevent these risks, another system known as "antivirus" was introduced. Similarly, technologies like AI and cloud computing have significantly improved the overall healthcare quality and clinical decision support system by enhancing data sharing, data availability, reducing cognitive load and so on; However, while doing so it also made it susceptible to data breaching. Government and FDA regulations can help immunizing healthcare industry based on the proposed framework. From a "system thinking" perspective a system that takes in feedback from its interacting sub-systems and a adjusts itself to adapt to the shifting paradigm and attain homeostasis is termed as an open system.

We have demonstrated the paradigm shift in the healthcare industry caused by the AI. This new paradigm has its own problems of cognitive bias among clinicians and data bias within AI algorithms, solutions to which is yet not readily available. Thus, to fill this gap, we analyzed the overall healthcare system and proposed a framework to address concerns regarding CDSS. Additionally, we believe that the impact of cognitive biases on clinical decisions can be minimized if the clinicians are made aware of their biases. AI in this situation cannot address the issues created by cognitive biases but is introduces a new bias – data bias. Further research should focus on the interaction between human bias and AI bias, will these two biases end up creating a resonance? Or will help CDSS become un-biased.